\magnification=\magstep1 
\font\bigbfont=cmbx10 scaled\magstep1
\font\bigifont=cmti10 scaled\magstep1
\font\bigrfont=cmr10 scaled\magstep1
\vsize = 23.5 truecm
\hsize = 15.5 truecm
\hoffset = .2truein
\baselineskip = 14 truept
\overfullrule = 0pt
\parskip = 3 truept
\def\frac#1#2{{#1\over#2}}

\nopagenumbers
%
\topinsert
\vskip 3.2 truecm
\endinsert
\centerline{\bigbfont BEYOND THE GROSS-PITAEVSKII EQUATION:}
\vskip 6 truept
\centerline{\bigbfont GROUND STATE AND LOW ENERGY EXCITATIONS}
\vskip 6 truept
\centerline{\bigbfont FOR TRAPPED BOSONS}
\vskip 20 truept
\centerline{\bigifont A. Polls}
\vskip 8 truept
\centerline{\bigrfont Departament d'Estructura i Constituents de la Mat\`eria,}
\vskip 2 truept
\centerline{\bigrfont Universitat de Barcelona} 
\vskip 2 truept
\centerline{\bigrfont Diagonal 647, Barcelona 08028, Spain} 
\vskip 14 truept
\centerline{\bigifont A. Fabrocini}
\vskip 8 truept
\centerline{\bigrfont Dipartimento di Fisica, Universit\`a di Pisa}
\vskip 2 truept
\centerline{\bigrfont and INFN, Sezione di Pisa, I-56100 Pisa, Italy} 
\vskip 1.8 truecm

\centerline{\bf 1.  INTRODUCTION}
\vskip 12 truept

The recent experimental realization
of Bose-Einstein condensation (BEC) of magnetically trapped alkali atoms has
generated a huge amount of experimental and theoretical activity, with 
over twenty experimental groups in the world that can produce
atomic condensates and more than one thousand articles
on the subject published by now. The present status of the 
field has been recently reviewed by F. Dalfovo et al.[1]. 

The history of BEC gets back to 1924, when S. Bose
derived the law for black-body radiation by considering the photons as 
a gas of identical particles. Immediately after, A. Einstein generalized
Bose's ideas to an ideal Bose gas and predicted that 
 the bosons would condensate in the lowest quantum state of
the system at sufficiently low temperature. 
It took seventy years of experimental efforts and thechnological progress
until  
the first atomic Bose condensate was achieved in 1995.
In this year, groups at the University of Colorado and at the
Massachusetts Institute of Technology (MIT)  using   a laser cooled
and magnetically trapped dilute gas of $^{87}$Rb [2] and $^{23}$Na  [3] atoms
respectively, were able to demonstrate  unambigously the occurrence of BEC.

In those early experiments,  the number of trapped atoms was relatively small 
 ($N\sim 10^3$ atoms) and  the transition temperature was 
around 100 nK. 
The magnetic trap
is well described by a harmonic  oscillator potential, usually with 
cylindrical symmetry. However, througout  this paper we 
will consider a spherical potential well confining the atoms.

In order to have quantum effects, i.e., wave behaviour, we 
need a de Broglie wave length $\lambda =(2 \pi \hbar^2 / m  T)^{1/2}$
 of the order of the distance
between the  atoms ($\rho \lambda^3 \sim 1$). On the other hand, the system
should be kept dilute, therefore the critical temperature 
will be extremely low, of the order of nanokelvins. 
 
Up to now, the 
experimental conditions were such that the atomic gas was very dilute, 
i.e., the average distance between the atoms is much larger than the range
of the interaction. As a consequence, the physics should be dominated by 
two-body collisions, generally well described in terms of the $s$-wave 
scattering length $a$. The case of a positive scattering length is equivalent
 to considering  a very dilute system of hard spheres, whose diameter coincides
with the scattering length itself. 

Typical scattering lengths  are 53 $\AA$ for $^{87}$Rb and 28 $\AA$ for 
$^{23}$Na. On the other hand, the size of the trap is defined by the 
harmonic oscillator length $a_{{\rm HO}}=(\hbar/m \omega)^{1/2}$
 which is of the 
order of $10^4 \AA$. The corresponding 
distance between the  energy levels associated with this potential well is 
around 4 nK.
  For those initial experiments,
a common $^{87}$Rb atom density in the trap was $\rho \sim 10^{12}-10^{14}$ 
atoms/cm$^3$ giving an average inter-atom distance $d \sim \rho^{-1/3} 
\sim 10^4 \AA$. Therefore, the effective atom size, defined by the
scattering length is usually small compared to both the trap size and the 
inter-atom distance. The crucial parameter that defines the 
condition of diluteness is $x=\rho  a^3$, which until very recently 
was kept rather small  (i.e., $x \sim 10^{-5}$). Under 
these conditions, the Gross-Pitaevskii equation [4], which assumes 
all the particles  in the condensate, seems the logical tool to 
study those systems.
 
The situation is somehow different in homogeneous liquid $^4$He. 
In this case,  BEC manifests itself as 
a macroscopic occupation of the zero momentum state,
 measured by the condensate fraction, i.e., the fraction of the 
total number of particles in this state. 
 However, there is  
 only indirect evidence for this macroscopic occupation. 
 Theoretical calculations and  the analysis of inelastic neutron 
scattering  data predict a condensate fraction  
 of $\sim 10$ $\%$ [5]. Such a large depletion is an indication that
$^4$He liquid is a strongly correlated system.

There are two ways to bring  $x$ outside the regime of validity 
of the mean field description. The first consists in 
increasing the density and the second in changing the effective size 
of the atoms. Recent  experiments have explored both possibilities.   
On one side they
have reached a very high number of atoms in the condensate, $\sim 10^8$,
 and on the other they have been able to change the scattering length of 
the atoms. This is  the case of a recent  experiment employing $^{85}$Rb, 
 where, by taking advantage of the presence of a Feshbach resonance at 
a magnetic field $B\sim 155$ Gauss, it was  possible to vary  
the scattering length  from negative to very high positive values.  
Under these conditions, effects beyond
the mean field approximation are expected to be observable [6,7].

 We will start by discussing a homogeneous system
of Bose hard spheres and using the results to determine the regime 
of validity of the Gross-Pitaevskii equation. 
We will then present an extension of the GP equation [8], still 
in the framework of mean field theories, that allows for giving 
a first estimate of the expected corrections to  the GP results
in these new scenarios. Finally, we will briefly analyze the effects on the  
low collective excitations of the system.  
\vskip 28 truept

\centerline{\bf 2.  Homogeneous hard-sphere Bose gas}
\vskip 12 truept

We consider a system of $N$ spinless bosons having mass $m$ and 
described by the many-body Hamiltonian:
$$ 
H= - \frac {\hbar^2}{2m} \sum_i \nabla_i^2 + \sum_{i<j} V(r_{ij}).
\eqno(1)
$$
The uniform system is studied in the thermodynamic limit,  
$N \rightarrow \infty $ and $\Omega  \rightarrow \infty$ keeping
the density, $\rho = N/\Omega$,  constant.  The hard-spheres potential
is  defined as $V(r<a) = \infty$ and $V(r>a)=0$.

Correlated Basis Functions (CBF) theory provides a very efficient way 
to handle the correlations induced by the interactions between 
the particles (for a review, see Ref. [9]). 
In its simplest version, one takes a Jastrow correlated wave function [10] 
$$ 
\Psi_J(1, ... ,N) = \prod_{i<j} f(r_{ij}), \eqno(2)
$$
where the Jastrow correlation function, $f(r)$, depends only on the
 interparticle distance. 
Once the trial function is defined, the variational principle ensures that, 
if we are capable to calculate the expectation value of the Hamiltonian,
$$
E_{{\rm CBF}} = \frac {\langle \Psi_J\mid H \mid \Psi_J\rangle}
{\langle \Psi_J \mid \Psi_J \rangle}, \eqno(3)
$$
then $E_{{\rm CBF}}$ is an upper bound to the true ground state energy.
The correlation function, $f(r)$,  is variationally determined 
by minimizing $E_{{\rm CBF}}$.

$E_{{\rm CBF}}$ may be calculated  once 
the two-body distribution function, $g(r)$, is known. 
In fact, the energy per particle can be written as
$$ 
e=  \frac {1}{2} \rho \int d^3 r~ g(r) \left [ V(r) - \frac {\hbar^2}{2 m}
\nabla^2 \ln f(r) \right ]. \eqno(4)
$$
In the particular case of hard spheres, the distribution function is strictly
zero for $r< a$ and the previous expression reduces just 
to the kinetic energy part
$$
e= - \frac {1}{2} \rho \int d^3 r~ g(r)  \frac {\hbar^2}{2 m}
\nabla^2 \ln f(r) . \eqno(5)
$$ 
$g(r)$ may be evaluated by using cluster expansion and Hypernetted Chain  
(HNC) theory. HNC is an integral equation method which allows for massive 
summations of the cluster diagrams associated with $g(r)$.
 
The optimal choice for the Jastrow factor would be the one satisfying the 
Euler equation $\delta E_{{\rm CBF}}/\delta f =0$. However, a 
less demanding and often effective approach consists in chosing a 
 parametrized functional form of $f(r)$ and in minimizing the energy with respect 
to the parameters.  We adopt here the correlation function minimizing 
the two--body cluster energy of a homogeneous Bose gas, with the 
healing conditions at a distance $d$ ($f(r\geq d)=1$ and $f'(d)=0$). 
$d$ is taken as a variational parameter. 
For the hard-spheres case, $f(r<a)=0$ and 
$f(r>a)=u(r)/r$, where $u(r)$ is the solution of the Schr\"odinger-like 
equation: $-u'' = K^2 u$. $f(r)$ has the form [11]
$$
f(r) = \frac {d}{r} \frac {\sin[K(r-a)]}{\sin[K(d-a)]}, \eqno(6)
$$
and the healing conditions are satisfied through  
the relation: $\cot [K(d-a)] =(Kd)^{-1}$.
\topinsert
\input psfig.sty
\centerline{\psfig{figure=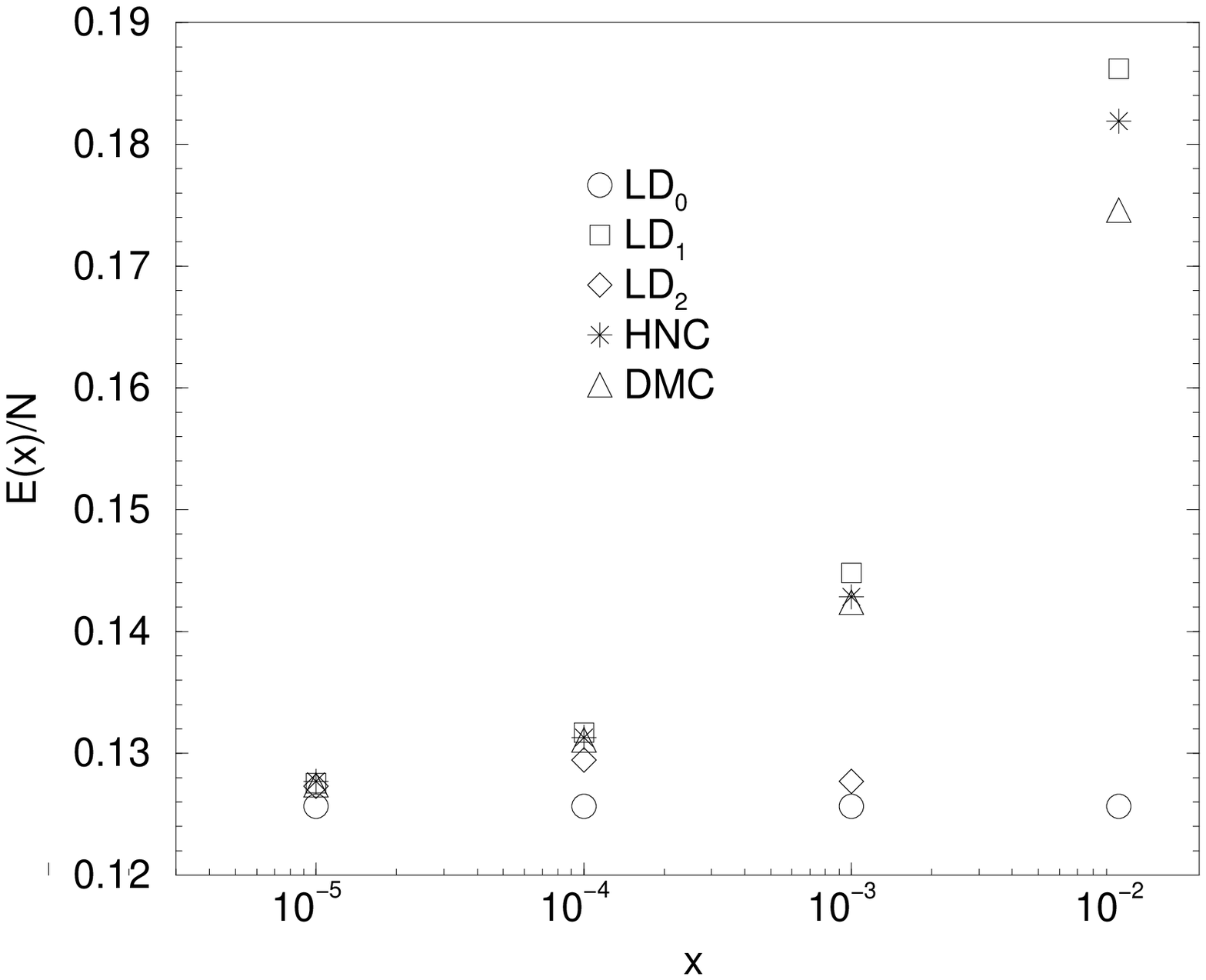,height=10truecm,width=15truecm,angle=0}}
\vskip -1 truecm
\hskip -1.5truecm

\noindent
{\bf Figure 1.} 
Energy per particle (in units of $\hbar^2/2m a^2$)
 for homogeneous hard
spheres in function of $x$. The symbols correspond to the low-density
expansion results obtained by keeping only the first term (LD$_0$) or by
adding the second (LD$_1$) and the third (LD$_2$) ones, and to the 
diffusion Monte Carlo (DMC) and HNC energies. The LD$_2$ energy at $x=10^{-2}$ 
is $E/N=0.07 $ and lies outside the frame. 
\vskip 12truept
\endinsert


An alternative calculation, based on perturbation theory in the 
expansion parrameter $x=\rho a^3$,  
 leads to the low-density expansion for the energy density[12]:
$$
\frac {E}{\Omega} = \frac {2 \pi \rho^2 a \hbar^2}{m} \left [ 1 +
\frac {128}{15} {\sqrt { \frac {\rho a^3}{\pi} }} +
8 \left ( \frac {4}{3} \pi - {\sqrt {3}} \right ) \rho a^3 ~\ln 
(\rho a^3) + O(\rho a^3) \right ] . \eqno(7)
$$ 
Up to these orders of the expansion, the details of the potential do not 
show up, and any potential with the same scattering length would give identical 
results. This universal behavior has  recently been checked by 
a diffusion Monte Carlo calculation (DMC)[13], which provided the 
exact solution of the many--body Schr\"odinger equation.

Fig. 1.  shows the energy per particle in 
units of $\hbar^2/2m a^2$ for
homogeneous hard spheres as a function of $x$. The energies have been 
multiplied by $10^{3(2,1)}$ at $x=10^{-5(-4,-3)}$, respectively. 
The figure compares the energies computed by 
retaining different terms  in expansion (7). The 
LD$_{0}$ values contain the first term only, whereas LD$_1$ and  
LD$_2$ are obtained by adding the second and third terms, respectively. 
The HNC results have been obtained disregarding the elementary diagrams 
(HNC/0) and using the correlation function of Eq.(6). The
DMC results correspond to diffusion Monte Carlo calculations [13].

The agreement  between the HNC/0 and the DMC  results 
is excellent in most of the wide range of densities considered. However, 
there is a 5\% disagreement at the highest $x$--value ($x=10^{-2}$). 
From this value on, the contribution
of the elementary diagrams  and a better optimization of 
the correlation should be probably taken into account. The LD$_0$ results
are only accurate at very low densities, while LD$_1$  gives  also a 
good representation of the exact DMC results.   
On the contrary, the addition of the logarithmic term spoils
the agreement  already at intermediate densities.

In the next section, we will describe the trapped bosons 
 by a local density approximation (LDA) employing the homogeneous gas results. 
 The local value of the parameter $x$ in the trap will 
give an idea of the differences that we can expect  by using the  
different energies reported in Fig. 1 as inputs to build
the energy functional. 
\vskip 28 truept

\centerline{\bf 2.  Ground state of trapped hard spheres}
\vskip 12 truept

The energy functional associated with the Gross-Pitaevskii theory is simply 
obtained in the local-density approximation by keeping only the first term
in the low density expansion (7):
$$
E_{{\rm GP}}[\Psi] = \int d{\bf r} \left [ \frac{\hbar^2}{2m}  
|{\bf \nabla} \Psi(r)|^2 + \frac {m}{2} \omega^2 r^2 |\Psi(r)|^2
+ \frac {2 \pi \hbar^2 a}{m} |\Psi(r)|^4 \right ], \eqno(8)
$$
where the wave function $\Psi(r)$, in which all the atoms 
belong to the condensate, is normalized to $N$.  
By a functional variation of $E_{{\rm GP}}[\Psi]$ 
one finds the Gross-Pitaevskii equation,
$$
\left [ - \frac{\hbar^2}{2m} \nabla^2 + \frac {m}{2} \omega^2 r^2 + 
\frac {4 \pi \hbar^2 a }{m} |\Psi(r)|^2 \right ] \Psi(r) = \mu 
\Psi(r), \eqno(9)
$$
where $\mu$ is the chemical potential. The GP equation has the form of a 
nonlinear stationary Schr\"odinger equation, and it has been solved for 
several types of traps using different numerical methods.

The next logical step, in the spirit of LDA, is to include into the energy
functional the next terms of the correlation energy expansion for the 
uniform system.
According to the behavior of the different terms, shown in Fig. 1, 
it seems clear that it is reasonable to consider only the first correction, 
$LD_1$. However, before 
proceeding further it is convenient to simplify the notation by expressing 
lengths and energies in harmonic oscillator (HO) units. 
The spatial coordinates, the energy, and the wave function are rescaled as 
${\bf r} =a_{{\rm HO}} {\bf \bar r}$,
$E=\hbar \omega \bar E$, and $\Psi(r)=(N/a_{{\rm HO}}^3)^{1/2} \Psi_1(\bar r)$, 
where $\Psi_1(\bar r)$ is normalized to unity.

Using these new variables and taking into account the second term
of the expansion, we obtain the modified Gross-Pitaevskii (MGP)
energy functional for the energy per atom, 
$\bar e_{{\rm MGP}}= \bar E_{{\rm MGP}}/N$,
$$
\bar e_{{\rm MGP}}[\Psi_1]= \int d {\bf \bar r} \left [ \frac {1}{2}  
|\nabla_{\bar r} \Psi_1|^2 + \frac {1}{2} \bar r^2 
|\Psi_1|^2 + 2\pi \bar a N |\Psi_1|^4 +
\frac {256 {\sqrt {\pi \bar a^{5} N^{3}}} }{15} |\Psi_1|^{5} 
\right ],\eqno(10)
$$
and  the  corresponding modified Gross-Pitaevskii equation,
$$
\left [ -\frac {1}{2} \nabla_{\bar r}^2 + \frac {1}{2} \bar r^2 + 
4 \pi \bar a N |\Psi_1(\bar r)|^2 + 
 {\sqrt {\pi \bar a^{5} N^{3}}} \frac {128}{3}
|\Psi_1(\bar r)|^3 \right ] \Psi_1(\bar r) = \mu_1 \Psi_1(\bar r),
\eqno(11)
$$
where $\bar a=a/a_{{\rm HO}}$ and $\bar \mu$ is the chemical potential 
in HO units.

In alternative, we  have also used as local correlation 
energy the one provided by the uniform system HNC results. 
This option has the advantage that
one is not limited to hard spheres, but, in principle, 
any type of potential for the two-body interaction can be considered. 
In this case, the local correlation energy 
$V_{{\rm corr}}^{LD}$, is given by
$$
V_{{\rm corr}}^{LD} = \frac {1}{N} \int d{\bf \bar r} \rho_1(\bar r) 
e_{{\rm HNC}}^{hom}
(\rho_1). \eqno(12)
$$
where $e_{{\rm HNC}}^{hom}(\rho_1)$ is the HNC homogeneous gas energy per particle at
density $\rho_1$. The minimization of the energy gives the
HNC correlated Hartree equation (CH$_{{\rm HNC}}$),
$$
\left [ - \frac{1}{2} \nabla_{\bar r}^2 + \frac {1}{2} \bar r^2 + 
\bar e_{\rm HNC}^{hom}(x_{{\rm loc}}) + x_{{\rm loc}}
 \frac {\partial \bar e_{\rm HNC}^{hom}(x_{{\rm loc}})}
{\partial x_{{\rm loc}}} \right ] \Psi_1(\bar r) = \mu_1 \Psi_1 (\bar r), 
\eqno(13) 
$$
where we have introduced the scaled unities and the local gas 
parameter, $x_{{\rm loc}}(\bar r) =\rho_1(\bar r) a^3 =
 N \bar a^3 |\Psi_1(\bar r)|^2$.

The GP, MGP and  CH$_{{\rm HNC}}$ equations have been solved by the 
steepest descent method for an  isotropic harmonic oscillator trap. 

Several relationships between the different contributions to the 
energy per atom or to the chemical potential exist, and are useful 
to check the accuracy of the numerical procedure.  
By direct integration of the GP equation, one finds 
$$
\bar \mu = \bar e_{{\rm kin}}+ \bar e_{{\rm HO}}+ 2 \bar e_{{\rm int}}^{(1)},
\eqno(14)
$$
where 
$\bar e_{{\rm kin}}=-\frac {1}{2} \int d^3 \bar r \Psi_1 \nabla^2 \Psi_1$, 
$\bar e_{{\rm HO}}=\frac {1}{2} \int d^3 \bar r |\Psi_1|^2 r^2 $ and 
$\bar e_{{\rm int}}^{(1)}=
\int d^3 \bar r 2 \pi a N |\Psi_1|^4$
 are the different terms
contributing to the total energy. Further relationships can be obtained by
means of the virial theorem,
$$
2 \bar e_{{\rm kin}}  - 2 \bar e_{{\rm HO}} + 3 \bar e_{{\rm int}}^{(1)} = 0.
\eqno(15)
$$

It is also important to notice that the dimensionless parameter characterizing
the effects of the interaction in the GP equation is given by $\bar a N$. 
This implies that one can get the same results with a proper rescaling
of the variables N and $\bar a$. 
As it can be seen by a simple inspection of the equations, the 
scaling property  is lost in the MGP approach. Moreover, 
 the relation between the different contributions to the chemical potential 
changes to
$$
\bar \mu = \bar e_{{\rm kin}}+  \bar e_{{\rm HO}}+ 2 \bar e_{{\rm int}}^{(1)}
+ \frac {5}{2} \bar e_{{\rm int}}^{(2)}, 
\eqno(16)
$$
where 
$$
\bar e_{{\rm int}}^{(2)}= \int d^3 \bar r ~\Psi_1(\bar r) \left [
 \frac {256}{15}{\sqrt {\pi \bar a^{5} N^{3}}}
 \Psi_1(\bar r)^3 \right ] \Psi_1 (\bar r).
\eqno(17) 
$$
In this case the relation implied  by the virial theorem is
$$
2 \bar e_{{\rm kin}}  - 2 \bar e_{{\rm HO}} + 3 \bar e_{{\rm int}}^{(1)}
 + \frac {9}{2} \bar e_{{\rm int}}^{(2)}= 0. 
\eqno(18)
$$

A simple approach, valid for large $N$ and 
loosely called the Thomas-Fermi (TF) approximation, 
 is obtained by neglecting  the kinetic
energy term in the GP equation. In the TF approximation 
it is possible to  derive simple analytical expressions [14], 
useful to make quick estimates of several quantities. 
For instance, 
$\bar \mu_{{\rm TF}} = 1/2 (15 \bar a N)^{2/5}$, 
while the  energy is related to the chemical
potential by  $\bar e_{{\rm TF}}=5 \bar \mu_{{\rm TF}} /7$. 
The local value of  $x= N \bar a^3 \rho_1(0)$ 
at the center
of the density distribution of the trapped bosons is given by 
$x_{{\rm TF}}(0)=\left ( 15^2 \bar a^{12} N^2 \right )^{1/5}/(8 \pi) $.

  
As can be seen in Table I, for large number of particles the TF and GP results are 
 practically identical and there is also a very good agreement
between the MGP and CH$_{{\rm HNC}}$ results. 
In the case $N=10^7$, the contributions to the energy for the GP(MGP) equations are: 
 $\bar e_{{\rm kin}}= 0.0294 (0.0292)$, $\bar e_{{\rm HO}}= 45.306 (46.57)$
, $\bar e_{{\rm int}}^{(1)}=30.184 (28.96)$ and $\bar e_{{\rm int}}=0 (1.379)$. 
The virial theorem is well fulfilled in both cases .  

\topinsert
\centerline{\bf{Table 1}}
\vskip 12 truept
\noindent
Chemical potential $\mu$, and ground state energy per particle $e$, of $N$
$^{87}$Rb atoms in an isotropic trap ($\omega/2 \pi = 77.78$Hz) in different
approaches. Energies are in units of $\hbar \omega$.
\vskip 24 truept

%
\newbox\hdbox%
\newcount\hdrows%
\newcount\multispancount%
\newcount\ncase%
\newcount\ncols
\newcount\nrows%
\newcount\nspan%
\newcount\ntemp%
\newdimen\hdsize%
\newdimen\newhdsize%
\newdimen\parasize%
\newdimen\spreadwidth%
\newdimen\thicksize%
\newdimen\thinsize%
\newdimen\tablewidth%
\newif\ifcentertables%
\newif\ifendsize%
\newif\iffirstrow%
\newif\iftableinfo%
\newtoks\dbt%
\newtoks\hdtks%
\newtoks\savetks%
\newtoks\tableLETtokens%
\newtoks\tabletokens%
\newtoks\widthspec%
%
%
\immediate\write15{%
CP SMSG GJMSINK TEXTABLE --> TABLE MACROS V. 851121 JOB = \jobname%
}%
%
%
\tableinfotrue%
\catcode`\@=11
%
%
\def\tstrut{\vrule height3.1ex depth1.2ex width0pt}%
\def\and{\char`\&}
\def\tablerule{\noalign{\hrule height\thinsize depth0pt}}%
\thicksize=1.5pt
\thinsize=0.6pt
\def\thickrule{\noalign{\hrule height\thicksize depth0pt}}%
\def\ctr#1{\hfil\ #1\hfil}%
%
%
%
%
\tablewidth=-\maxdimen%
\spreadwidth=-\maxdimen%
\def\tabskipglue{0pt plus 1fil minus 1fil}%
%
%
\centertablestrue%
%
%
%
%
\parasize=4in%
\gdef\ARGS{########}
\gdef\headerARGS{####}
\def\@mpersand{&}
{\catcode`\|=13
\gdef\letbarzero{\let|0}
\gdef\letbartab{\def|{&&}}%
\gdef\letvbbar{\let\vb|}%
}
{\catcode`\&=4
\def\ampskip{&\omit\hfil&}
\catcode`\&=13
\let&0
\xdef\letampskip{\def&{\ampskip}}%
\gdef\letnovbamp{\let\novb&\let\tab&}
}
\def\begintable{
   \begingroup%
   \catcode`\|=13\letbartab\letvbbar%
   \catcode`\&=13\letampskip\letnovbamp%
   \def\multispan##1{
      \omit \mscount##1%
      \multiply\mscount\tw@\advance\mscount\m@ne%
      \loop\ifnum\mscount>\@ne \sp@n\repeat%
   }
   \def\|{%
      &\omit\widevline&%
   }%
   \ruledtable
}
\long\def\ruledtable#1\endtable{%
%
%
%
   \offinterlineskip
   \tabskip 0pt
   \def\widevline{\vrule width\thicksize}
   \def\endrow{\@mpersand\omit\hfil\crnorm\@mpersand}%
   \def\crthick{\@mpersand\crnorm\thickrule\@mpersand}%
   \def\crnorule{\@mpersand\crnorm\@mpersand}%
   \let\nr=\crnorule
   \def\endtable{\@mpersand\crnorm\thickrule}%
   \let\crnorm=\cr
%
%
   \edef\cr{\@mpersand\crnorm\tablerule\@mpersand}%
   \the\tableLETtokens
%
%
   \tabletokens={&#1}
%
%
   \countROWS\tabletokens\into\nrows%
   \countCOLS\tabletokens\into\ncols%
%
%
   \advance\ncols by -1%
   \divide\ncols by 2%
   \advance\nrows by 1%
%
%
   \iftableinfo %
      \immediate\write16{[Nrows=\the\nrows, Ncols=\the\ncols]}%
   \fi%
%
%
   \ifcentertables
      \ifhmode \par\fi
      \line{
      \hss
   \else %
      \hbox{%
   \fi
      \vbox{%
         \makePREAMBLE{\the\ncols}
         \edef\next{\preamble}
         \let\preamble=\next
         \makeTABLE{\preamble}{\tabletokens}
      }
      \ifcentertables \hss}\else }\fi
   \endgroup
   \tablewidth=-\maxdimen
   \spreadwidth=-\maxdimen
}
\def\makeTABLE#1#2{
   {
   \let\ifmath0
   \let\header0
   \let\multispan0
%
%
   \ncase=0%
   \ifdim\tablewidth>-\maxdimen \ncase=1\fi%
   \ifdim\spreadwidth>-\maxdimen \ncase=2\fi%
   \relax
%
   \ifcase\ncase %
      \widthspec={}%
   \or %
      \widthspec=\expandafter{\expandafter t\expandafter o%
                 \the\tablewidth}%
   \else %
      \widthspec=\expandafter{\expandafter s\expandafter p\expandafter r%
                 \expandafter e\expandafter a\expandafter d%
                 \the\spreadwidth}%
   \fi %
   \xdef\next{
      \halign\the\widthspec{%
      #1
      \noalign{\hrule height\thicksize depth0pt}
      \the#2\endtable
%
      }
   }
   }
   \next
}
\def\makePREAMBLE#1{
   \ncols=#1
   \begingroup
   \let\ARGS=0
   \edef\xtp{\widevline\ARGS\tabskip\tabskipglue%
   &\ctr{\ARGS}\tstrut}
   \advance\ncols by -1
   \loop
      \ifnum\ncols>0 %
      \advance\ncols by -1%
      \edef\xtp{\xtp&\vrule width\thinsize\ARGS&\ctr{\ARGS}}%
   \repeat
   \xdef\preamble{\xtp&\widevline\ARGS\tabskip0pt%
   \crnorm}
   \endgroup
}
\def\countROWS#1\into#2{
   \let\countREGISTER=#2%
   \countREGISTER=0%
   \expandafter\ROWcount\the#1\endcount%
}%
\def\ROWcount{%
   \afterassignment\subROWcount\let\next= %
}%
\def\subROWcount{%
   \ifx\next\endcount %
      \let\next=\relax%
   \else%
      \ncase=0%
      \ifx\next\cr %
         \global\advance\countREGISTER by 1%
         \ncase=0%
      \fi%
      \ifx\next\endrow %
         \global\advance\countREGISTER by 1%
         \ncase=0%
      \fi%
      \ifx\next\crthick %
         \global\advance\countREGISTER by 1%
         \ncase=0%
      \fi%
      \ifx\next\crnorule %
         \global\advance\countREGISTER by 1%
         \ncase=0%
      \fi%
      \ifx\next\header %
         \ncase=1%
      \fi%
      \relax%
      \ifcase\ncase %
         \let\next\ROWcount%
      \or %
         \let\next\argROWskip%
      \else %
      \fi%
   \fi%
   \next%
}
\def\counthdROWS#1\into#2{%
\dvr{10}%
   \let\countREGISTER=#2%
   \countREGISTER=0%
\dvr{11}%
\dvr{13}%
   \expandafter\hdROWcount\the#1\endcount%
\dvr{12}%
}%
\def\hdROWcount{%
   \afterassignment\subhdROWcount\let\next= %
}%
\def\subhdROWcount{%
   \ifx\next\endcount %
      \let\next=\relax%
   \else%
      \ncase=0%
      \ifx\next\cr %
         \global\advance\countREGISTER by 1%
         \ncase=0%
      \fi%
      \ifx\next\endrow %
         \global\advance\countREGISTER by 1%
         \ncase=0%
      \fi%
      \ifx\next\crthick %
         \global\advance\countREGISTER by 1%
         \ncase=0%
      \fi%
      \ifx\next\crnorule %
         \global\advance\countREGISTER by 1%
         \ncase=0%
      \fi%
      \ifx\next\header %
         \ncase=1%
      \fi%
\relax%
      \ifcase\ncase %
         \let\next\hdROWcount%
      \or%
         \let\next\arghdROWskip%
      \else %
      \fi%
   \fi%
   \next%
}%
{\catcode`\|=13\letbartab
\gdef\countCOLS#1\into#2{%
   \let\countREGISTER=#2%
   \global\countREGISTER=0%
   \global\multispancount=0%
   \global\firstrowtrue
   \expandafter\COLcount\the#1\endcount%
   \global\advance\countREGISTER by 3%
   \global\advance\countREGISTER by -\multispancount
}%
\gdef\COLcount{%
   \afterassignment\subCOLcount\let\next= %
}%
{\catcode`\&=13%
\gdef\subCOLcount{%
   \ifx\next\endcount %
      \let\next=\relax%
   \else%
      \ncase=0%
      \iffirstrow
         \ifx\next& %
            \global\advance\countREGISTER by 2%
            \ncase=0%
         \fi%
         \ifx\next\span %
            \global\advance\countREGISTER by 1%
            \ncase=0%
         \fi%
         \ifx\next| %
            \global\advance\countREGISTER by 2%
            \ncase=0%
         \fi
         \ifx\next\|
            \global\advance\countREGISTER by 2%
            \ncase=0%
         \fi
         \ifx\next\multispan
            \ncase=1%
            \global\advance\multispancount by 1%
         \fi
         \ifx\next\header
            \ncase=2%
         \fi
         \ifx\next\cr       \global\firstrowfalse \fi
         \ifx\next\endrow   \global\firstrowfalse \fi
         \ifx\next\crthick  \global\firstrowfalse \fi
         \ifx\next\crnorule \global\firstrowfalse \fi
      \fi
\relax
      \ifcase\ncase %
         \let\next\COLcount%
      \or %
         \let\next\spancount%
      \or %
         \let\next\argCOLskip%
      \else %
      \fi %
   \fi%
   \next%
}%
\gdef\argROWskip#1{%
   \let\next\ROWcount \next%
}
\gdef\arghdROWskip#1{%
   \let\next\ROWcount \next%
}
\gdef\argCOLskip#1{%
   \let\next\COLcount \next%
}
}
}
\def\spancount#1{
   \nspan=#1\multiply\nspan by 2\advance\nspan by -1%
   \global\advance \countREGISTER by \nspan
   \let\next\COLcount \next}%
\def\dvr#1{\relax}%
\def\header#1{%
\dvr{1}{\let\cr=\@mpersand%
\hdtks={#1}%
\counthdROWS\hdtks\into\hdrows%
\advance\hdrows by 1%
\ifnum\hdrows=0 \hdrows=1 \fi%
\dvr{5}\makehdPREAMBLE{\the\hdrows}%
\dvr{6}\getHDdimen{#1}%
{\parindent=0pt\hsize=\hdsize{\let\ifmath0%
\xdef\next{\valign{\headerpreamble #1\crnorm}}}\dvr{7}\next\dvr{8}%
}%
}\dvr{2}}
\def\makehdPREAMBLE#1{
\dvr{3}%
\hdrows=#1
{
\let\headerARGS=0%
\let\cr=\crnorm%
\edef\xtp{\vfil\hfil\hbox{\headerARGS}\hfil\vfil}%
\advance\hdrows by -1
\loop
\ifnum\hdrows>0%
\advance\hdrows by -1%
\edef\xtp{\xtp&\vfil\hfil\hbox{\headerARGS}\hfil\vfil}%
\repeat%
\xdef\headerpreamble{\xtp\crcr}%
}
\dvr{4}}
\def\getHDdimen#1{%
\hdsize=0pt%
\getsize#1\cr\end\cr%
}
\def\getsize#1\cr{%
\endsizefalse\savetks={#1}%
\expandafter\lookend\the\savetks\cr%
\relax \ifendsize \let\next\relax \else%
\setbox\hdbox=\hbox{#1}\newhdsize=1.0\wd\hdbox%
\ifdim\newhdsize>\hdsize \hdsize=\newhdsize \fi%
\let\next\getsize \fi%
\next%
}%
\def\lookend{\afterassignment\sublookend\let\looknext= }%
\def\sublookend{\relax%
\ifx\looknext\cr %
\let\looknext\relax \else %
   \relax
   \ifx\looknext\end \global\endsizetrue \fi%
   \let\looknext=\lookend%
    \fi \looknext%
}%
%
%
\def\tablelet#1{%
   \tableLETtokens=\expandafter{\the\tableLETtokens #1}%
}%
\catcode`\@=12

\nrows=5
\ncols=9
\begintable
 \| & &$\mu$ & | & & $e$ & \crthick 
N \|  TF  & GP & MGP & CH$_{{\rm HNC}}$  
| TF  & GP  & MGP & CH$_{{\rm HNC}}$  \crthick
$10^5$      \|  ~16.75&~16.85&~16.99&~16.94 | ~11.96&~12.10&12.19&~12.20~ \crnorule 
$10^6$      \|  ~42.07&~42.12&~42.69&~42.53 | ~30.05&~30.12&~30.48&~30.48~\crnorule 
$10^7$      \|  ~105.68&~105.70&~107.97&~107.20 | ~75.49&~75.52&~76.94&~76.85~  
\endtable
\vskip 14 truept
\endinsert

\noindent


By changing the number of trapped atoms in the range of 
the experimental availability, the average values  of $x$  are  such that the 
corrections  to the GP equation are kept small and of the order of 2$\%$
in the case of $N=10^7$ atoms. However, the recent experiments, where
 the scattering length can be largely manipulated, open the door to explore
 higher values of $x$. In fact, in order to vary $x$, it is much more efficient 
to change the scattering length than
 the number of atoms. Experimental results are  available  
for  $^{85}$Rb,  whose scattering length is modulated by its 
  Feshbach resonance. The number of trapped atoms is $N \sim 10^4$ 
and the trap is anisotropic. In order to estimate the 
corrections induced by the MGP equation, we have considered an
isotropic trap characterized by the frequency $\omega/2\pi =10$ Hz,  
 corresponding to an average of the cylindrical trap frequencies used 
in the experiment. 
The $\omega$ value is smaller than that employed for
$^{87}$Rb and therefore $a_{{\rm HO}}$  is larger. We take
$\bar a=0.1228$, which is in the range considered by the experiments.
It corresponds to $a = 8000 a_0$, where $a_0$ is the Bohr radius of
the hydrogen atom.  In this case, $x_{{\rm TF}}(0)=0.03 $, which is 
just beyond the range of the points plotted in 
 Fig.1. The energies per atom  turn out to be: 
$\bar e_{{\rm GP}}=18.25$ and $\bar e_{{\rm MGP}}=21.85$. 
For the chemical potential we have:
 $\bar \mu_{{\rm GP}}=25.48$ and $ \bar \mu_{{\rm MGP}}=31.09$.
 As a consequence of the use of the MGP equation, 
the corrections are of the order of 20 $\%$. The actual cylindrical 
trap is currently under study [15]. 
 
\vskip 28 truept
\centerline{\bf 4.  COLLECTIVE EXCITATIONS}
\vskip 12 truept

Information on the excitation spectrum of a system are  contained in the
dynamic structure function, which, for a given excitation operator $F$, 
is given by
$$
S_F(E) = \sum_n |\langle n |F|0\rangle |^2 \delta(E-(E_n-E_0)). \eqno(19)
$$
In the case of inelastic neutron scattering ( for instance
against liquid $^4$He) the operator $F$ corresponds to the density fluctuation
 operator, $F=\sum_j e^{i{\bf q r}_j}$, and the inclusive inelastic 
cross section is proportional to the dynamic structure function. 

A useful tool to anlyze $S_F(E)$ is provided by the sum rules 
approach, extensively used in quantum many-body systems 
both in the context of nuclear [16] and condensed matter physics, in particular
to analyze the excitation spectrum of quantum liquids [5]. The sum rules
establish rigorous links among the energy momenta of $S_F(E)$ 
and ground state properties. 
The energy weighted integrals are defined as:
$$
m_k = \int_0^{\infty} E^k S_F(E) dE ,\eqno(20)
$$
and $m_k$ can  be calculated, without an explicit knowledge of $S_F(E)$, 
as a ground state expectation value of certain combinations
of commutators of the excitation operator $F$ and the Hamiltonian.
 So, quantities like $m_{k+1}/m_k$ or $(m_{k+2}/m_k)^{1/2}$ supply
 rigorous upper
bounds to the energy of the lowest excited state that can be connected to 
the ground state through the operator $F$. The upper bounds are very 
reliable when the excited state is highly collective, i.e., when the 
strength distribution is almost exhausted by a single mode. We will 
concentrate here in the case of the compressional modes, 
or monopole excitations, 
whose  associated excitation operator is $F=\sum_i^N r_i^2$. 

In the present case, we will study  the upper bound provided by
 $(m_3/m_1)^{1/2}$. Using the definition of the dynamic structure function 
and the completness of the 
eigenstates $|n\rangle$, the moments $m_1$ and $m_3$ can be expressed as
$$
m_1 = {{1}\over {2}} \langle 0 |[F^{\dagger}, [H,F]]|0 \rangle \eqno(21)
$$
and 
$$
m_3= {{1}\over {2}} \langle 0 | [[F^{\dagger},H],[H,[H,F]]]|0\rangle~. \eqno(22)
$$
By explicitly calculating the commutators, $m_1$ is expressed in terms
of the mean square radius, 
$$
m_1 = 2 {{\hbar^2}\over {m}} N \langle 0 | r^2 | 0\rangle ~. \eqno(23)
$$
Obviously, we get the same expression of $m_1$ for both GP and MGP equations.

In the case of the monopole excitation, is more efficient to calculate
$m_3$ as
$$
m_3 ={{1}\over {2}} \left ({{2 \hbar^2}\over {m}} \right )^2 
{{d^2 E(\lambda)}\over {d \lambda^2}} |_{\lambda=1}~, \eqno(24)
$$
where $\lambda$ is the parameter of the scaling transformation 
$\rho(r) \rightarrow \lambda^3 \rho(\lambda r) $.  In fact, for the 
GP equation, the energy for the scaled density is
$$
E(\rho,\lambda) = N \left [ \lambda^2 e_{kin}(\rho) +
{{e_{HO}}\over {\lambda^2}}+ \lambda^3 e_{int}^{(1)}~\right ]~. \eqno(25) 
$$
The condition $dE(\rho,\lambda)/d \lambda =0$  at $\lambda=1$
satisfies the virial theorem.  $m_3$ is given by
$$
m_3 =  \left ( {{2 \hbar^2}\over {m}} \right )^2 
\hbar^2 N  e_{HO}
\left [5 - {{e_{kin}}\over {e_{HO}}}\right ]~.   \eqno(26)
$$
Finally
$$
 E_{ex}^{GP} \sim {\sqrt {{m_3}\over {m_1}}} = \hbar \omega
 \left [5 - {{e_{kin}}\over 
{e_{HO}}} \right ] ^{1/2}~. \eqno(27)
$$
By a similar procedure for the MGP equation we get
$$
E_{ex}^{MGP} = \hbar \omega 
\left [ 5 - {{e_{kin}}\over {e_{HO}}} + {{27}\over {8}}
{{e_{int}^{(2)}}\over {e_{HO}}} \right ]^{1/2}~. \eqno(28)
$$

In the $^{85}$Rb case the estimates 
of the monopole excitation energies (in HO units) are 2.23 and 2.38 
for the GP and MGP, respectively. So, the MGP correction to the 
excitation energy is about  7$\%$. This correction 
is smaller than that to the energy itself. The whole spectrum 
is shifted but the separation between the excitation energy levels
 is less affected.
 
\vskip 28 truept

\centerline{\bf 4.  CONCLUSIONS}
\vskip 12 truept

In conclusion, we find that the MGP equation induces corrections 
of $20\%$ in the ground state properties of the condensate, 
when the conditions of the recent experiments for 
$^{85}$Rb are considered ($x\sim 10^{-2}$). MGP is still 
a mean field theory, since it tries to incorporate correlation 
effects into the average single particle potential, and  
it cannot predict the depletion of the condensate. However, we believe
that the estimates of the energy, chemical potential, and 
density profile are surely indicative and can still be accurate. 
Moreover, it is legitimate at these densities to question the 
use of a simplified interaction, given in terms of hard spheres. 
 In any case, it is clear that fully microscopic calculations, 
which might address the many--body wave function and take explicitly 
into account the depletion of the condensate, are urgently required [17].  
\vskip 24 truept

\centerline{\bf ACKNOWLEDGMENTS}
\vskip 12 truept

We thank X. Vi\~nas for helpful discussions. 
This research was supported by
DGICYT (Spain) Grant No.PB98-1247, the program SGR98-11 from 
Generalitat de Catalunya and the agreement CICYT (Spain)-INFN (Italy).

\vskip 24 truept

\centerline{\bf REFERENCES}
\vskip 12 truept

\item{[1]}
F. Dalfovo, {\it et al.}, 
{\it Rev.Mod.Phys.} {\bf 71}, 463 (1999).
\item{[2]}
M.H. Anderson, {\it et al.},  
{\it Science} {\bf 269}, 198 (1995).
\item{[3]}
K.B. Davis {\it et al.}, 
{\it Phys.Rev.Lett.} {\bf 75}, 1687 (1995). 
\item{[4]}
L.P. Pitaevskii, {\it Sov.Phys.JETP} {\bf 13}, 451 (1961);
E.P. Gross, {\it Nuovo Cimento} {\bf 20}, 454 (1961).
\item{[5]}
H.R. Glyde, {\it Excitations in Liquid and Solid Helium},
Clarendon Press, Oxford 1994. 
\item{[6]}
S.L. Cornish {\it et al.},  
{\it Phys.Rev.Lett.} {\bf 85}, 1795 (2000). 
\item{[7]}
 {\it Search and Discovery,} in {\it Phys.Today}, 
17 (August 2000).  
\item{[8]}
A.~ Fabrocini and A.~Polls, {\it Phys.Rev.} {\bf A60}, 2319 (1999). 
\item{[9]}
S.~Fantoni and A.~Fabrocini, in {\it Microscopic Quantum Many
Body Theories and Their Applications}, J. Navarro and A. Polls eds., 
Lecture Notes in Physics Vol.150 (Springer-Verlag, Berlin, 1998)119.
\item{[10]}
R.~Jastrow, {\it Phys.Rev.} {\bf 98}, 1479 (1955).
\item{[11]}
V.R.~Pandharipande and K.E.~Schmidt,{\it Phys.Rev.} {\bf A15}, 2486 (1977).
\item{[12]}
A.L.~Fetter and J.D.~Walecka, {\it Quantum Theory of 
Many-Particle Systems} (McGraw-Hill, New York, 1971).
\item{[13]}
S.~Giorgini, J.~Boronat and J.~Casulleras, {\it Phys.Rev.} {\bf A60}, 5129
 (1999).
\item{[14]}
M.~Edwards and K.~Burnett, {\it Phys.Rev.} {\bf A51}, 1382 (1995).
\item{[15]}
A.~Fabrocini and A.~Polls, in preparation.
\item{[16]}
E.~Lipparini and S.~Stringari, {\it Phys.Rep.} {\bf 175}, 103 (1989).
\item{[17]} 
J.L.~Dubois and H.R.~Glyde, {\it cond-mat}/0008368 and contribution to 
this volume. 
\end